# THE EFFECT OF Mn INTERSTITIALS ON THE LATTICE PARAMETER OF $Ga_{1-x}Mn_xAs$

I.Kuryliszyn-Kudelska[a,b], J.Z.Domagala[a], T.Wojtowicz[a,b], X.Liu[b], E. Łusakowska[a], W.Dobrowolski[a], J.K.Furdyna[b]

[a]*Institute of Physics, Polish Academy of Sciences, Warsaw, PL-02-668, Poland*

[b]*Department of Physics, University of Notre Dame, Notre Dame, IN 46556, USA*

**ABSTRACT**

Structural investigation of as-grown as well as annealed $Ga_{1-x}Mn_xAs$ epilayers was carried out using high resolution X-ray diffraction (XRD) measurements for a wide range of Mn concentrations ($0.027 \leq x \leq 0.083$), with special attention on how the interstitial Mn atoms ($Mn_I$) influence the lattice parameter of this material. We observe a distinct decrease of the lattice parameter after low temperature annealing of $Ga_{1-x}Mn_xAs$, which is known to reduce the $Mn_I$ concentration. The reciprocal space maps measured for all the investigated samples showed that the $Ga_{1-x}Mn_xAs$ layers are fully strained - i.e., they remain pseudomorphic to the GaAs (001) substrate - for the entire thickness of the samples used (in the present case over 100 nm). In all cases studied the XRD measurements revealed high crystalline perfection of both as-grown as well as annealed $Ga_{1-x}Mn_xAs$ epilayers.

## I. INTRODUCTION

Ferromagnetic $Ga_{1-x}Mn_xAs$ is the subject of intense research due to its interesting physical properties, its high Curie temperature ($T_C$ > 100K), as well as its possible spin-electronics applications [1]. Recently, experimental studies revealed that the Curie temperature of $Ga_{1-x}Mn_xAs$ epilayers can be further increased by post-growth heat treatment (low temperature annealing) [2,3,4]. Our previous papers [4,5] indicated that low temperature (LT) annealing leads to a rearrangement of the Mn ions in the $Ga_{1-x}Mn_xAs$ host lattice. Specifically, channeling Rutherford backscattering (c-RBS) and channeling particle-induced X-ray emission (c-PIXE) experiments revealed that such LT annealing leads to the removal of Mn atoms from interstitial positions [5].

One should note that the interstitial $Mn_I$ atoms are not only relatively mobile, but they are double donors, and are thus positively charged. As they diffuse through the lattice, they are thus expected to be attracted to the negatively-charged substitutional Mn acceptors $Mn_{Ga}$ to form $Mn_{Ga}$-$Mn_I$ pairs. Theoretical calculations [6] showed that the $Mn_I$ atoms do not contribute to the ferromagnetic coupling mediated by free holes and, furthermore, that the $Mn_I$-$Mn_{Ga}$ pairs are coupled antiferromagnetically through superexchange interaction. Thus the increase of the hole concentration (measured by the electrochemical capacitance voltage profiling method, ECV), the increase of the Curie temperature, and the increase in the saturation magnetization observed on annealed samples all serve to corroborate that LT annealing results in the removal of a significant fraction of Mn interstitial atoms in $Ga_{1-x}Mn_xAs$.

It was experimentally established that the lattice constant of $Ga_{1-x}Mn_xAs$ layers increases with the increase of Mn concentration [7]. Recently, first-principles theoretical calculations [8] have predicted that the presence of Mn interstitials atoms can be the reason of for the observed expansion of the lattice constant of $Ga_{1-x}Mn_xAs$. Furthermore, it is well known that the presence of arsenic antisites ($As_{Ga}$) in low temperature (LT-GaAs) also leads to an increase in the lattice constant. The $As_{Ga}$ defects are also expected to be present in $Ga_{1-x}Mn_xAs$ layers, since these are grown at low temperatures similar to those used for growing LT-GaAs. It has been shown that the lattice constant of $Ga_{1-x}Mn_xAs$ additionally depends in a sensitive way on the growth conditions, quite possibly due to excess As incorporation, the degree of which may in itself depend on the presence of Mn in the system [9,10].

Measurements of the lattice constant *a* have often been used as a method of determining the Mn concentration *x* in $Ga_{1-x}Mn_xAs$. While there clearly exists a *phenomenological* correlation between *a* and *x*, based on the remarks just made it is now clear that this correlation is quite complex, and is not really understood. For example, it was recently reported that epilayers of $Ga_{1-x}Mn_xAs$ *with the same composition* but prepared using different growth parameters have quite different lattice constants [10]. Moreover, it was suggested that $Ga_{1-x}Mn_xAs$ is an example of a system that does not obey Vegard's law in the traditional sense of a linear variation between the two end-point compounds, i.e., between zinc blende GaAs and (hypothetical) zinc blende MnAs [10]. In order to contribute to resolving this issue, our goal in this paper is to explore specifically the influence of Mn interstitials atoms on the lattice constant of $Ga_{1-x}Mn_xAs$ which, as noted in ref. [8], is expected to strongly affect the lattice parameter. Here the decrease of the $Mn_I$ concentration with annealing provides an extremely valuable handle for examining the effect of $Mn_I$ on the lattice constant, and for comparing this effect with theoretical predictions. Specifically, the aim of this work is to perform systematic structural studies for a series of as-grown and annealed $Ga_{1-x}Mn_xAs$ epilayers in a wide range of Mn concentration, in order to assess how the removal of $Mn_I$ in this alloy affects the lattice constant.

## II. SAMPLE PREPARATION AND EXPERIMENTAL RESULTS

The $Ga_{1-x}Mn_xAs$ epilayers were grown using low temperature (LT) molecular beam epitaxy. Semi-insulating "epiready" (001) GaAs wafers were used as substrates. A buffer of GaAs was first grown at the substrate temperature of $600^0$C. The substrate was then cooled down to a temperature in the range of 265 to $270^0$C and a thin layer of LT-GaAs was deposited, followed by deposition of the $Ga_{1-x}Mn_xAs$ layer. The $Ga_{1-x}Mn_xAs$ layers were typically grown to a thickness in the range between 100 and 150 nm. The estimated values of the sample thicknesses from the RHEED oscillations as well as from the x-ray interference (pendeloesung) fringes observed in high resolution XRD measurements are shown in Table 1.

After removal from the MBE chamber, the as-grown samples were cleaved into a number of pieces for systematic annealing experiments. The annealing was found to be optimal (i.e., the highest Curie temperatures were obtained) when carried out at temperatures in the range between 260 and $290^0$C, under a fixed flow of $N_2$ gas of 1.5 SCFH (standard cubic feet per hour) for 1 hour. Additional details of the annealing procedure have been described elsewhere [4].

The Mn concentrations in the three $Ga_{1-x}Mn_xAs$ specimens used in this investigation were $x$ = 0.027, 0.062, and 0.083, determined as discussed in Appendix 1.

Comparison of the zero-field resistivity and SQUID magnetization measurements carried out on these samples before and after annealing all showed large increases in conductivity, in the Curie temperature, and in saturation magnetization of the samples annealed under the optimal conditions (i.e., for one hour at $T \approx 280^{o}C$). For example, for the sample with $x$ = 0.083 the Curie temperature (estimated from the temperature $T_\rho$ at which the zero-field resistivity shows a cusp) shifts from 88K for the as-grown sample to 127K after annealing, for $x$ = 0.061 $T_\rho$ shifts from 67 to 101K, and for the $Ga_{1-x}Mn_xAs$ layer with the lowest Mn concentration $T_\rho$ increases from 53.2 to 62K.

High resolution X-ray diffraction (XRD) studies were performed using a Philips X'Pert-MRD diffractometer equipped with a parabolic X-ray mirror, a four-bounce Ge 220 monochromator at the incident beam, and a three-bounce Ge analyzer at the diffracted beam. The XRD measurements revealed high crystalline perfection of the $Ga_{1-x}Mn_xAs$ layers both before and after they were annealed. Figure 1 shows the $\omega/2\theta$ scans obtained for the symmetric (004) reflection for the as-grown samples with $x$ = 0.027, 0.062 and 0.083. Figure 2 shows $\omega/2\theta$ scan for the sample with $x$ = 0.083 before and after annealing. For both as-grown and annealed samples the profiles of $\omega/2\theta$ scan exhibit clear interference fringes, attesting to the high structural perfection of the layers. These oscillations also provide a very direct method for determining the $Ga_{1-x}Mn_xAs$ layer thickness. Note that the values of the layer thickness obtained by this method (listed in Table 1) agree rather well with those obtained from RHEED oscillations, thus providing an added measure of internal consistency of the experiments under discussion. The good agreement of the experimental values of full width at half maximum (FWHM) observed for the (004) Bragg reflections (from 170 to 180 arcsec) with the FWHM values obtained from the simulated curves (~ 187 arcsec) shown in Figure 3 also indicate that the crystalline quality of the $Ga_{1-x}Mn_xAs$ layers is rather high. Finally, the lack of asymmetry in the profiles indicates the absence of detectable strain gradients within the entire thickness of the $Ga_{1-x}Mn_xAs$ film.

Additional information about the structure of the layers before and after annealing can be obtained from the reciprocal lattice maps shown in Figs. 4 and 5 for the (004) and (224) reflections, respectively. It is clearly seen from these figures that annealing does not alter the crystalline quality of the $Ga_{1-x}Mn_xAs$ layers. The very narrow $Q_x$-direction peaks corresponding to the $Ga_{1-x}Mn_xAs$ layers (unchanged by the annealing process), along with the presence of the interference fringes due to multiple reflections within the $Ga_{1-x}Mn_xAs$ layer, indicate a sharp interface between the $Ga_{1-x}Mn_xAs$ layer and the GaAs substrate. This, together with the relative sharpness of the $Ga_{1-x}Mn_xAs$ peak, suggests that the Mn content is uniform (negligible gradient in $x$) throughout the epilayer. The fact that the value of $Q_x$ in the asymmetric reciprocal lattice maps (see Fig. 5) is the same for the $Ga_{1-x}Mn_xAs$ layer and for the GaAs substrate also reveal that the $Ga_{1-x}Mn_xAs$ films (as-grown as well as annealed) are fully strained to the (100) GaAs substrate (i.e., fully pseudomorphic), with no detectable relaxation throughout the thickness of the film. The cross shown in Fig. 5 indicates the position where the peak from the hypothetical *relaxed* $Ga_{1-x}Mn_xAs$ would occur on the (224) reciprocal space map, thus serving to illustrate the degree of tetragonal distortion *uniformly* experienced by the $Ga_{1-x}Mn_xAs$ alloy along the growth direction.

Measurements of the (004) Bragg reflections allow us to calculate the lattice parameters perpendicular to the layer plane ($a_\perp$) for all samples studied. The combination of this and the measurements of the asymmetric (224) Bragg reflections are then used to determine the in-plane lattice parameter ($a_{\|}$). The values of relaxed layer lattice parameter $a_{relax}$ (calculated from the measured values of $a_{\|}$ and $a_{\perp}$), and the calculated values of the relaxed mismatch $(a_{relax}-a_s)/a_s$ before and after annealing (where $a_s$ is the lattice parameter of the GaAs substrate), are shown in Table 1. The lattice parameters $a_{relax}$ for relaxed $Ga_{1-x}Mn_xAs$ were obtained using the relation

$$a_{relax} = (a_{\perp}+2ba_{\|})/(1+2b), \qquad (1)$$

where $b = C_{11}/(C_{11}+2C_{12})$. Here $C_{11}$ and $C_{12}$ are elastic constants for GaAs ($C_{11}$ = 11.82x10$^{10}$ Pa, $C_{12}$ = 5.326x10$^{10}$ Pa [11]). As seen from the reciprocal space maps for the (224) reflection shown in Fig. 5, $a_{\|}$ is essentially identical to the lattice parameter of GaAs.

Analogous results were obtained for the samples with lower Mn content, $x$ = 0.062 and $x$ = 0.027. As in the case of $x$ = 0.083, both the $\omega/2\theta$ scans and the reciprocal lattice maps for the symmetric (004) and asymmetric (224) reflections indicated high crystalline perfection of the samples, and systematically revealed a clear decrease of the perpendicular lattice parameter after annealing. Similar trends have also been observed in other laboratories [12]. The results for $a_{\perp}$, $a_{\|}$, $a_{relax}$, and for the value of the relaxed lattice mismatch for all samples before and after annealing are collected in Table 1.

In addition to the above experiments we carried out surface studies of the as-grown and the annealed samples using atomic force microscope (AFM). Comparison of AFM images taken at different positions on the samples did not reveal any changes of the surface image after the annealing process.

## III. DISCUSSION

The primary result of this work is the observation that the lattice parameter decreases when interstitials are removed from the alloy. As noted earlier, this effect has been foreseen by Mašek *et al.* [8], and we can now compare our experimental results with their predictions. We know from earlier studies by c-RBS and c-PIXE [5] that in $Ga_{1-x}Mn_xAs$ epilayers similar to our $x = 0.083$ sample the concentration of $Mn_I$ before annealing was approximately 14% of the total Mn content, and after annealing was reduced to 7% of the total Mn [5]. We will therefore use our results obtained on the $x = 0.083$ sample for an illustrative calculation. Since we ascribe $x = 0.083$ to *substitutional* Mn (see Appendix 1), using the arguments discussed in the Appendix we have estimated the atomic fraction of $Mn_I$ ($x_{int}$) in this sample as 0.014 in the as-grown material. As noted above, after annealing $x_{int}$ decreases by about a factor of 2, for a change in interstitial concentration estimated at $\Delta x_{int} = 0.007$.

The calculation of Masek *et al.* indicate that the relaxed lattice parameter of $Ga_{1-x}Mn_xAs$ has the following form (in Å):

$$a = a_o + 0.02x_{sub} + 1.05x_{int} + 0.69y, \qquad (2)$$

where $a_o$ is the lattice parameter of GaAs, $x_{sub}$ and $x_{int}$ are the concentrations of substitutional and interstitial Mn, and $y$ is the concentration of As antisites $As_{Ga}$. It is safe to assume that after low-temperature annealing $x_{sub}$ and $y$ remain unchanged (the temperature $T < 300^oC$ is too low to "kick out" these atoms from the crystallographic Ga sites), so that the only significant effect of such annealing is the removal of $Mn_I$. Assuming $x_{sub}$ and $y$ to remain constant, and using the values of $a_o$ before and after annealing from Table 1 (essentially identical to $a_{\|}$), Eq. (1) gives a change of the interstitial concentration $\Delta x_{int}$ = 0.004. This is smaller, but of the same order of magnitude as the change in $x_{int}$ estimated from the RBS/PIXE experiments. Applying a similar analysis to the remaining lattice parameter changes listed in Table 1, we consistently get lower values from Eq. (2) for the decrease in the $Mn_I$ concentration (by a similar factor of about 2) than those obtained if one makes the (admittedly very rough) assumption that the value of $x_{int}$ is reduced from 14 to 7% of the total Mn content in all samples. Given the number of assumptions involved, the lack of a better agreement between Eq. (2) and our estimates of the changes in $x_{int}$ are of course not surprising.

More important is the fact that the observed trend is quite systematic: not only is the lattice parameter consistently lower for annealed samples, but the degree by which it is lower is proportional to the Mn concentration, as can be clearly seen from Fig. 6, where we have plotted the relaxed lattice constant before and after annealing as a function of the substitutional Mn concentration $x$. Two interesting features emerge from this figure. First, when one extrapolates the as-grown and annealed data points to $x$ = 1.0, one obtains, respectively, the values of 5.90 and 5.86Å for the hypothetical zinc blende MnAs. It is interesting that a theoretical calculation of $a$ for zinc blende MnAs using covalent radii $r_c$ (for As $r_c$ = 1.225Å; for Mn $r_c$ = 1.326Å) [13] gives a value of 5.89Å. While the agreement between this value and the extrapolations in Fig. 6 may be somewhat coincidental, this is probably the reason why the increase of $a$ with $x$, along with the use of Vegard's law, has been rather widely accepted for $Ga_{1-x}Mn_xAs$, and has been initially interpreted as the effect of substitutional Mn in this alloy.

Figure 6 also reveals another interesting feature. Note that the amount of decrease of $a$ after annealing appears to be proportional to $x$, suggesting that the annealing-induced drop in the concentration of $Mn_I$ increases with Mn concentration. Since we know from the c-PIXE results already cited that annealing reduces the $Mn_I$ concentration by roughly a factor of two [5], this would suggest that the difference between the lattice parameter of as-grown material and of $Ga_{1-x}Mn_xAs$ in which there are no Mn interstitials would be approximately twice as large as that seen in Fig. 6 between the as-grown and annealed cases. Extending this logic to $x = 1.0$, we obtain an estimate of 5.83Å for the lattice parameter of zinc blende MnAs *with only substitutional Mn*. This is of course in disagreement with Ref. [8], in which it is predicted that the effect of substitutional Mn on the lattice constant of $Ga_{1-x}Mn_xAs$ is practically negligible (in stark contradiction with the estimated values obtained by using covalent radii). This points to an interesting physical insight: if one assumes zinc blende $Ga_{1-x}Mn_xAs$ with only substitutional Mn, one must take into account that every Mn produces an uncompensated hole, and for MnAs (or even $Ga_{1-x}Mn_xAs$ with a large value of $x$) one automatically has a metal. It is likely that the Coulomb interaction between the hole gas and the positive ions experience a Coulomb attraction, exactly as in the case of a metallic bond, which causes a contraction of the lattice parameter. One would assume that this effect is present in the first-principles calculations discussed in Ref. [8] (although it may be over-estimated in the calculations). The tendency for $a$ to decrease further with decreasing concentration of $Mn_I$, as signaled by the results plotted in Fig. 6, may thus be a qualitative indication of the physical processes implicitly taken into account in the first-principles calculation of Mašek *et al.* [8]

We have already noted that our data indicated the samples to be uniformly strained. The degree of strain, defined as $(a_{relax} - a_s)/a_s$ (where $a_s$ is the lattice parameter of the underlying substrate) is also listed in the table. As expected, the degree of strain in the specimens increases with the Mn content; but the strain is *reduced* by the annealing process.

## APPENDIX 1: METHOD OF DETERMINING THE Mn CONCENTRATION *x*

The Mn concentrations in the three $Ga_{1-x}Mn_xAs$ specimens used in this investigation were $x$ = 0.027, 0.062, and 0.083. Although the knowledge of *absolute values* of $x$ are not essential for the present paper, it is nevertheless important to define the manner in which these concentration were established. This is because it has been amply shown that the value of $a$ for a given $x$ depends sensitively on growth conditions [9, 10], thus raising some doubt on its usefulness as a tool for determining $x$.

In our case the concentrations 0.083 and 0.062 were determined from the change in the rate of RHEED oscillations observed during growth (under excess As pressure) after the Mn shutter was opened. An example of such data is shown in Fig.7. Note the rate of growth measured by RHEED oscillations is *in terms of atomic layers per second*. After the Mn shutter is opened the growth rate increases in proportion to precisely that fraction of the Mn flux required for completion of atomic layers as the growth proceeds. We can thus assume that the change in the rate of RHEED oscillations provides a measure of the concentration of *substitutional Mn cations* $Mn_{Ga}$, since *only these ions* are required to form successive atomic layers. This interpretation of RHEED oscillations has been additionally confirmed by systematic channeled particle-induced X-ray emission (c-PIXE) measurements [5]. We note that c-PIXE measurements are extremely valuable in this context, since they can distinguish between the respective contributions from Mn ions in substitutional, interstitial, and "random" positions (i.e., those in the form of random precipitates, such as MnAs inclusions) to the total Mn concentration. The c-PIXE results have consistently indicated that in samples comparable to those used here the total Mn concentration (i.e., substitutional, interstitial, and random) were always higher by 15 to 20% than the values obtained from RHEED in as-grown samples, thus reinforcing. For example, PIXE measurements on an as-grown sample with *x* determined by RHEED to be 0.08 ±0.02 indicated that the total Mn content in that sample was 0.092, of which 0.072 was in substitutional positions, 0.013 in the form of interstitials, and 0.007 occurred as random precipitates (i.e., in the ratio $x_{tot}$**:**$x_{sub}$**:**$x_{int}$**:**$x_{random}$ ≈ 1.0**:**0.78**:**0.14**:**0.08). We have used this scaling to estimate the concentration of Mn interstitials in the illustrative calculation in the preceding section applied to our $x = 0.083$ sample.

Despite the uncertainty surrounding the value of *x* obtained from the lattice parameter, it is well established that for samples grown under identical conditions (i.e., on the same day and with the same temperature settings) the lattice parameter does increase approximately linearly with *x* [9,10]. We have therefore used the Mn concentration values established by RHEED for samples with high Mn content ($x = 0.062$ and 0.083) to calibrate the dependence of *a* on *x*, and have then used this *a* vs. *x* relation to obtain the Mn concentration for our most dilute sample ($x = 0.027$), where the change in the rate of RHEED oscillations is too small to be reliable.

We note parenthetically that the interpretation of RHEED oscillations as a measured of the concentration of subsitutionally-incorporated Mn is clearly corroborated by our experiments on co-doping $Ga_{1-x}Mn_xAs$ with Be. It has been established that the incorporation of Mn in $Ga_{1-x}Mn_xAs$ is limited by the Fermi level $E_F$ [5]. When we grew GaAs under Be flux that is sufficient to reach the limiting value of $E_F$, our attempts at simultaneously incorporating Mn resulted only in forming Mn interstitials [14]. Under these circumstances we clearly saw that the rate of RHEED oscillations *did not change* when the Mn shutter was opened during the growth, consistent with the interpretation that when Mn does not go into substitutional positions, it does not affect the rate of RHEED oscillations.

**ACKNOWLEDGMENTS:**

We thank G. Schott for sharing with us her unpublished XRD data. The work in the Institute of Physics of the Polish Academy of Sciences in 2003 was supported by a Grant from the Polish State Committee for Scientific Research, and at the University of Notre Dame by the National Science Foundation Grant DMR02-45227.

**FIGURE CAPTIONS:**

**Fig 1.** The $\omega/2\theta$ scan for the symmetric (004) Bragg reflection for as-grown samples with $x$ = 0.027, 0.062, and 0.083.

**Fig 2.** The $\omega/2\theta$ scan for the symmetric (004) Bragg reflection for as-grown and annealed sample with $x$ = 0.083.

**Fig 3.** Comparison of the measured $\omega/2\theta$ scan for the symmetric (004) Bragg reflection (dashed curve) with simulation results (solid curve) for the annealed sample with $x$ = 0.083. Apart from the good agreement *per se*, the simulation data provide a very precise measure of the thickness of the $Ga_{1-x}Mn_xAs$ film.

**Fig 4**. Reciprocal space maps of the symmetric (004) reflection for the **(a)** as-grown and **(b)** annealed sample with $x$ = 0.083. ($Q_x$ and $Q_z$ represent reciprocal space vectors: $Q_x$ is in the direction <110> parallel to the surface, $Q_z$ is in the direction perpendicular to the surface, both given in units of $\lambda/2d$, where $\lambda$ = 0.15406 nm and $d$ denotes the interplanar spacing for the (004) planes).

**Fig 5**. Reciprocal space maps of the asymmetric (224) reflection for the **(a)** as-grown and **(b)** annealed sample with $x$ = 0.083. ($Q_x$ and $Q_z$ represent reciprocal space vectors: $Q_x$ is in the direction <110> parallel to the surface, $Q_z$ is in the direction perpendicular to the surface, both given in units of $\lambda/2d$, where $\lambda$ = 0.15406 nm and $d$ denotes the interplanar spacing for the (224) planes).

**Fig 6.** Relaxed lattice parameter of $Ga_{1-x}Mn_xAs$ plotted as a function of the Mn concentration $x$ for as-grown and annealed samples.

**Fig 7.** RHEED oscillations observed during the growth of a $Ga_{1-x}Mn_xAs$ film with $x$ = 0.062. The first 7 periods correspond to LT-GaAs growth. The "jump" in the signal occurs at the point when the Mn shutter has been opened and the rate of oscillations increased.

**TABLE CAPTIONS:**

**Table1.** The measured values of perpendicular to the layer plane lattice parameter($a_{\perp}$), in-plane lattice parameter ($a_{\|}$), the calculated values of the relaxed mismatch $(a_{relax}\text{-}a_s)/a_s$, and thickness of the GaMnAs epilayers determined from both RHEED oscilations and XRD measurements before and after annealing.

| Sample | $a_{\|}$ [Å] | $a_{\perp}$ [Å] | $a_{relaxed}$ [Å] | Δa/a [ppm] | d [nm] XRD | d [nm] RHEED |
|---|---|---|---|---|---|---|
| **0.027 as-grown** | **5.65348** | **5.66941** | **5.661243** | **1373** | **122** | **131** |
| **0.027 annealed** | **5.65348** | **5.66829** | **5.660697** | **1277** | **123** | **-** |
| **0.062 as-grown** | **5.65348** | **5.68431** | **5.668505** | **2658** | **140** | **149** |
| **0.062 annealed** | **5.65348** | **5.68049** | **5.666643** | **2328** | **136** | **-** |
| **0.083 as-grown** | **5.65348** | **5.6953** | **5.67386** | **3605** | **98** | **105** |
| **0.083 annealed** | **5.65348** | **5.68783** | **5.67022** | **2961** | **101** | **-** |

**Table 1**

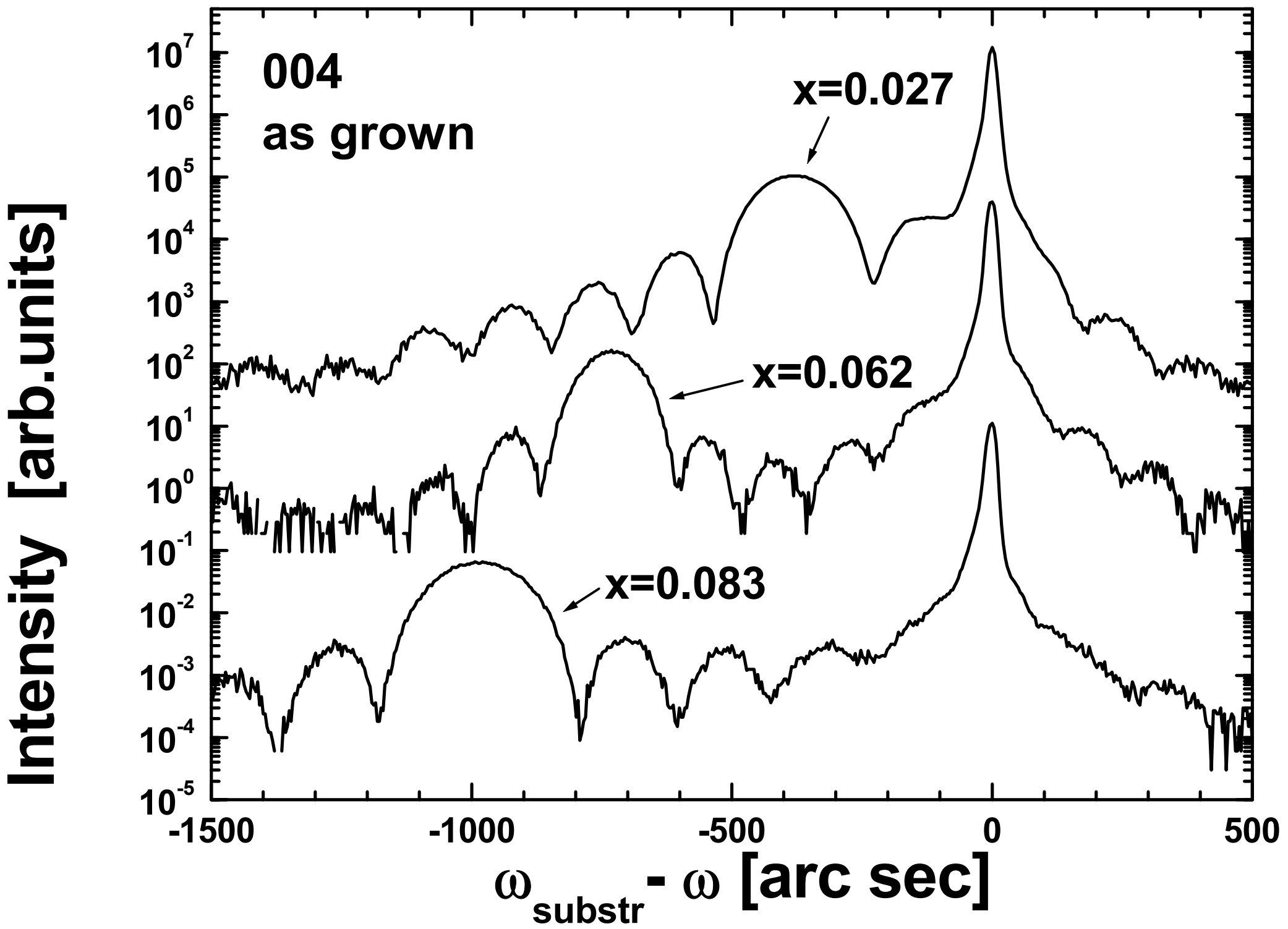

Fig.1

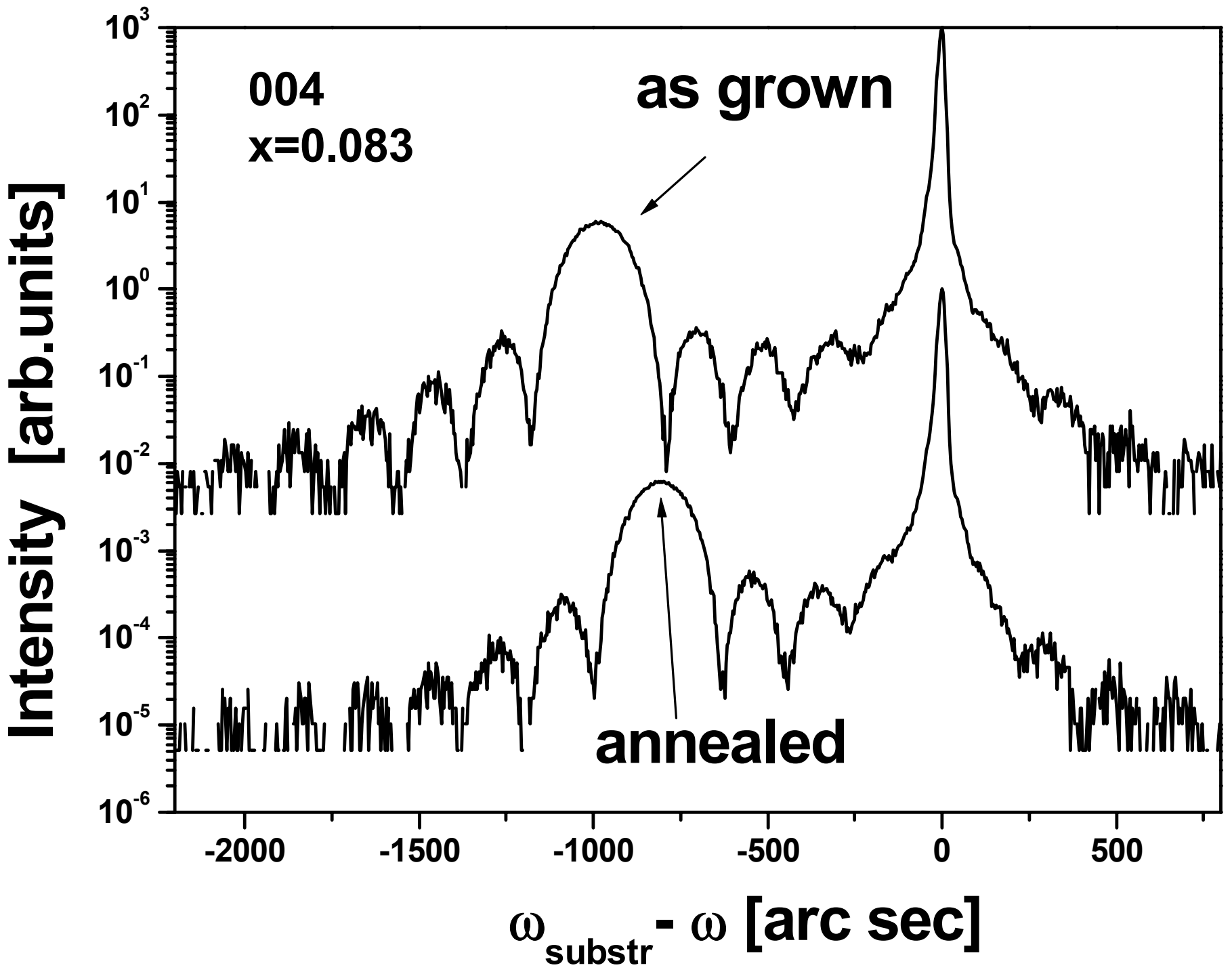

Fig.2

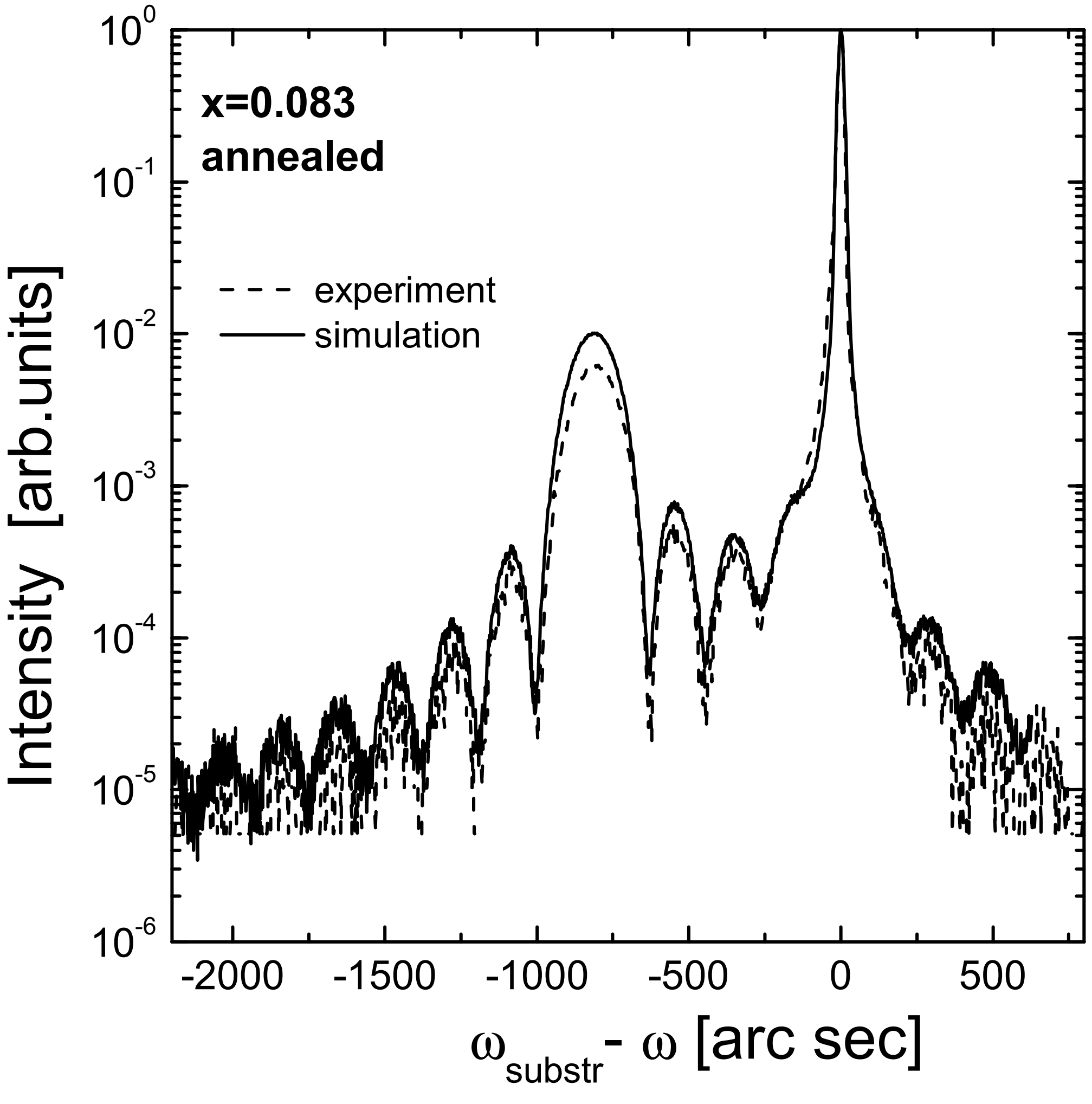

Fig. 3

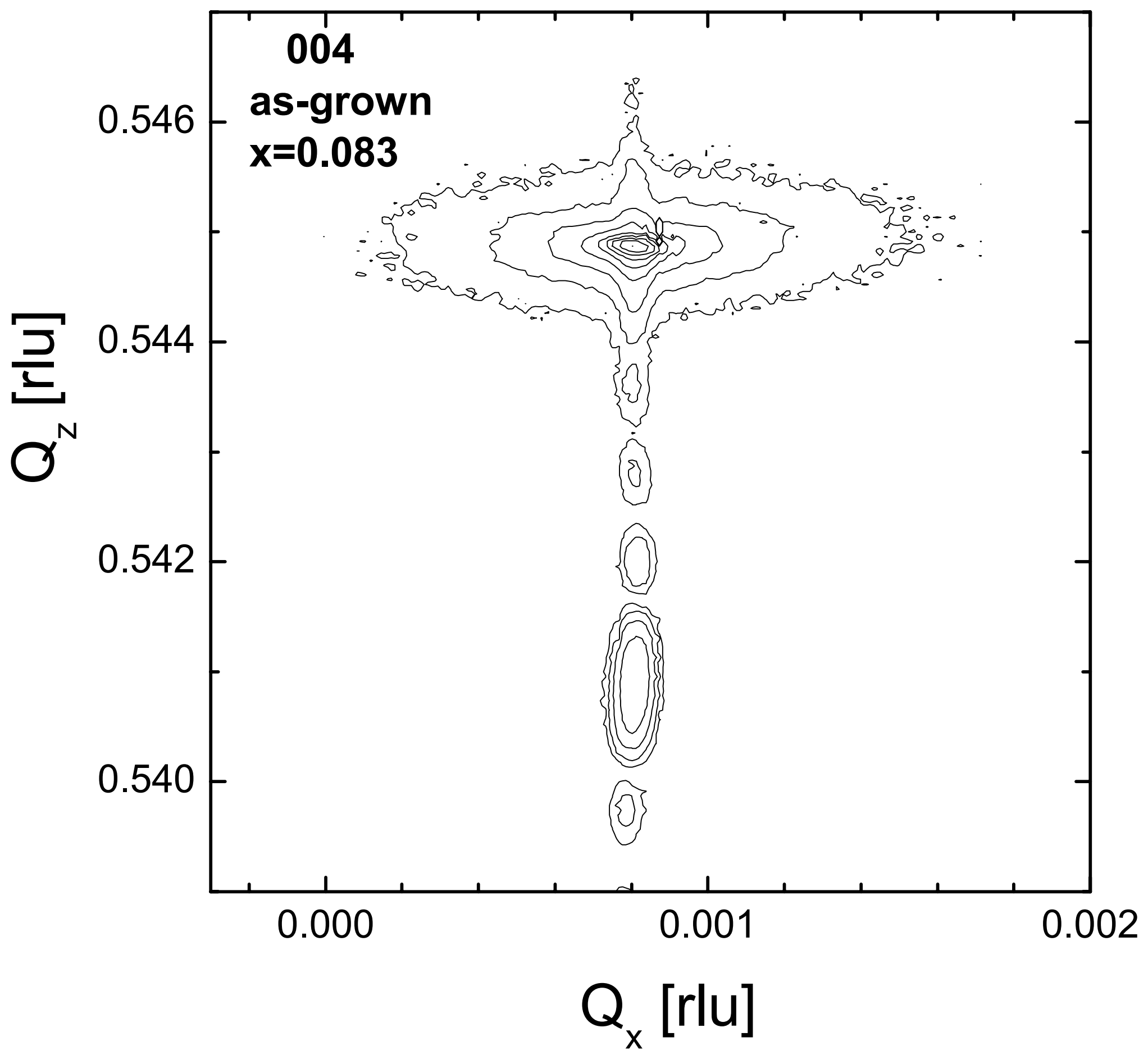

Fig. 4a

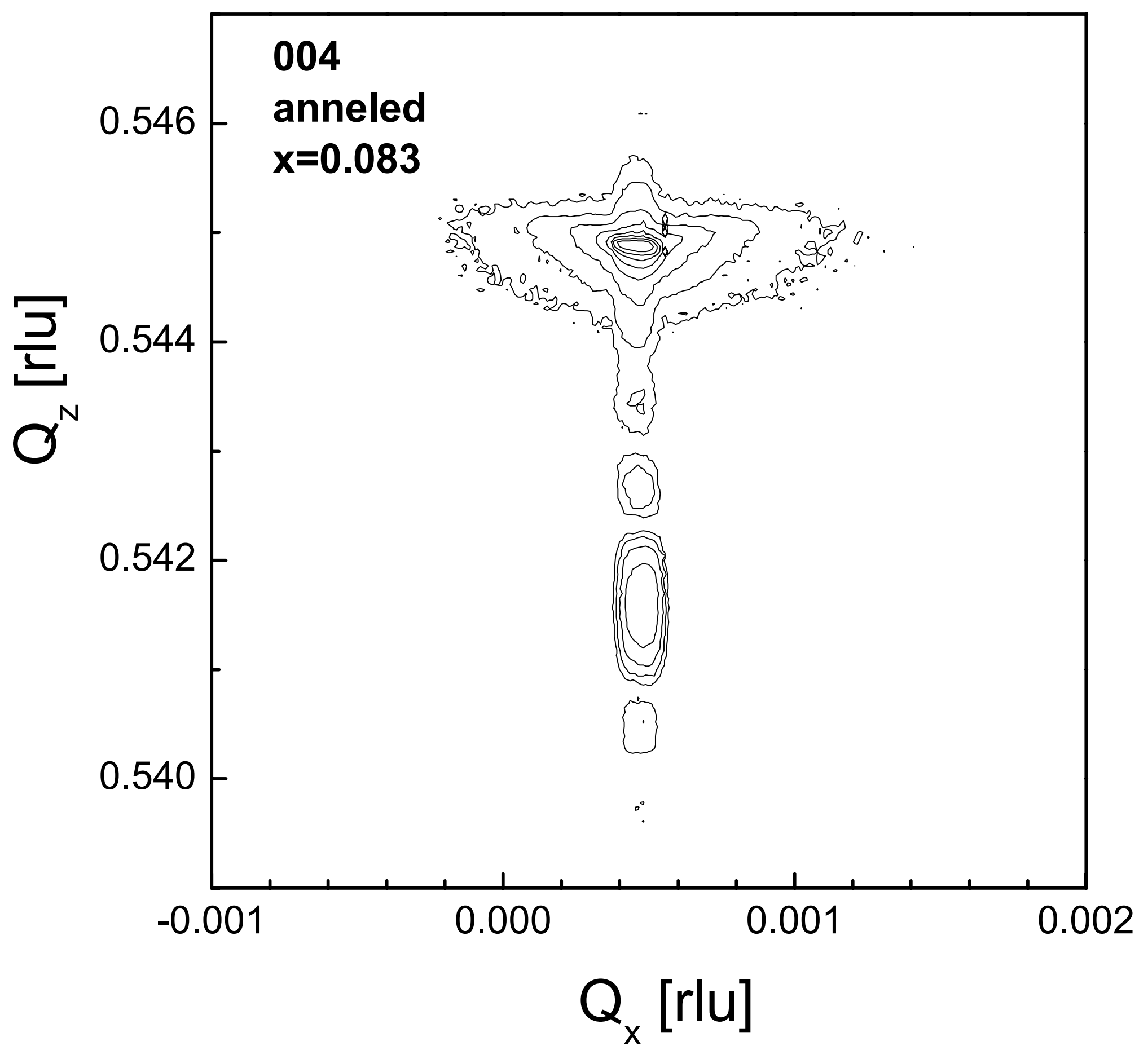

Fig. 4b

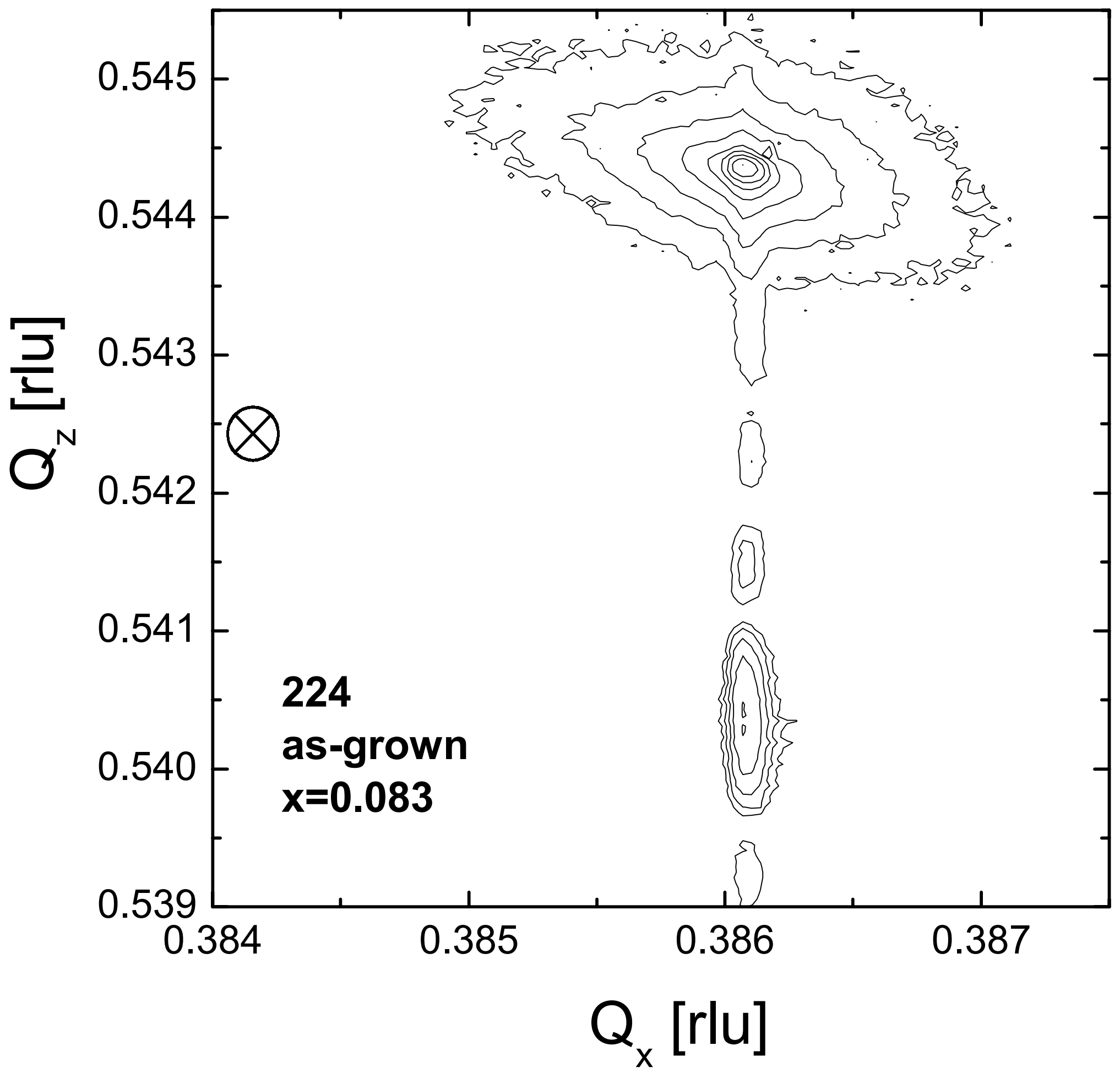

Fig. 5a

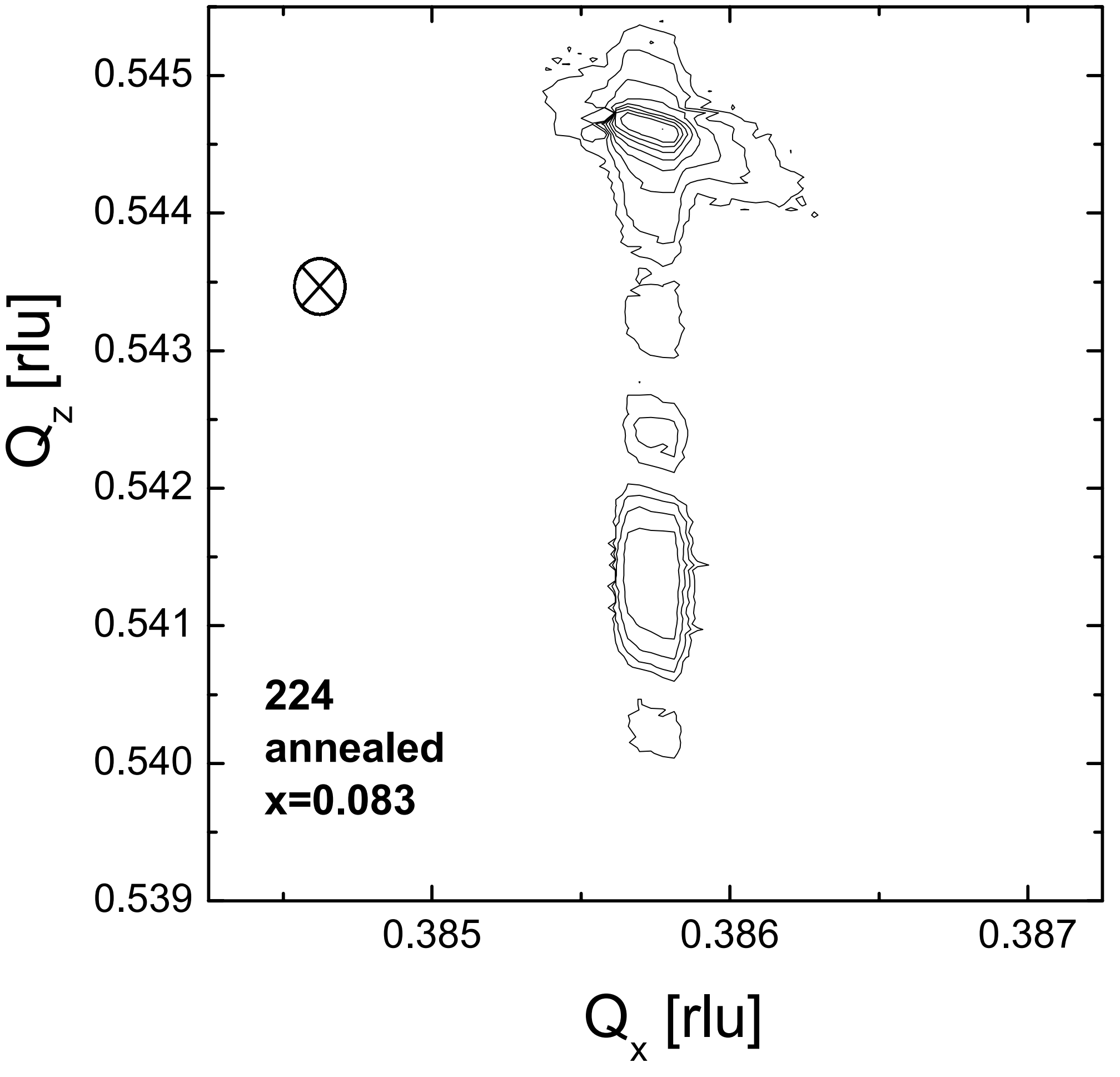

Fig. 5b

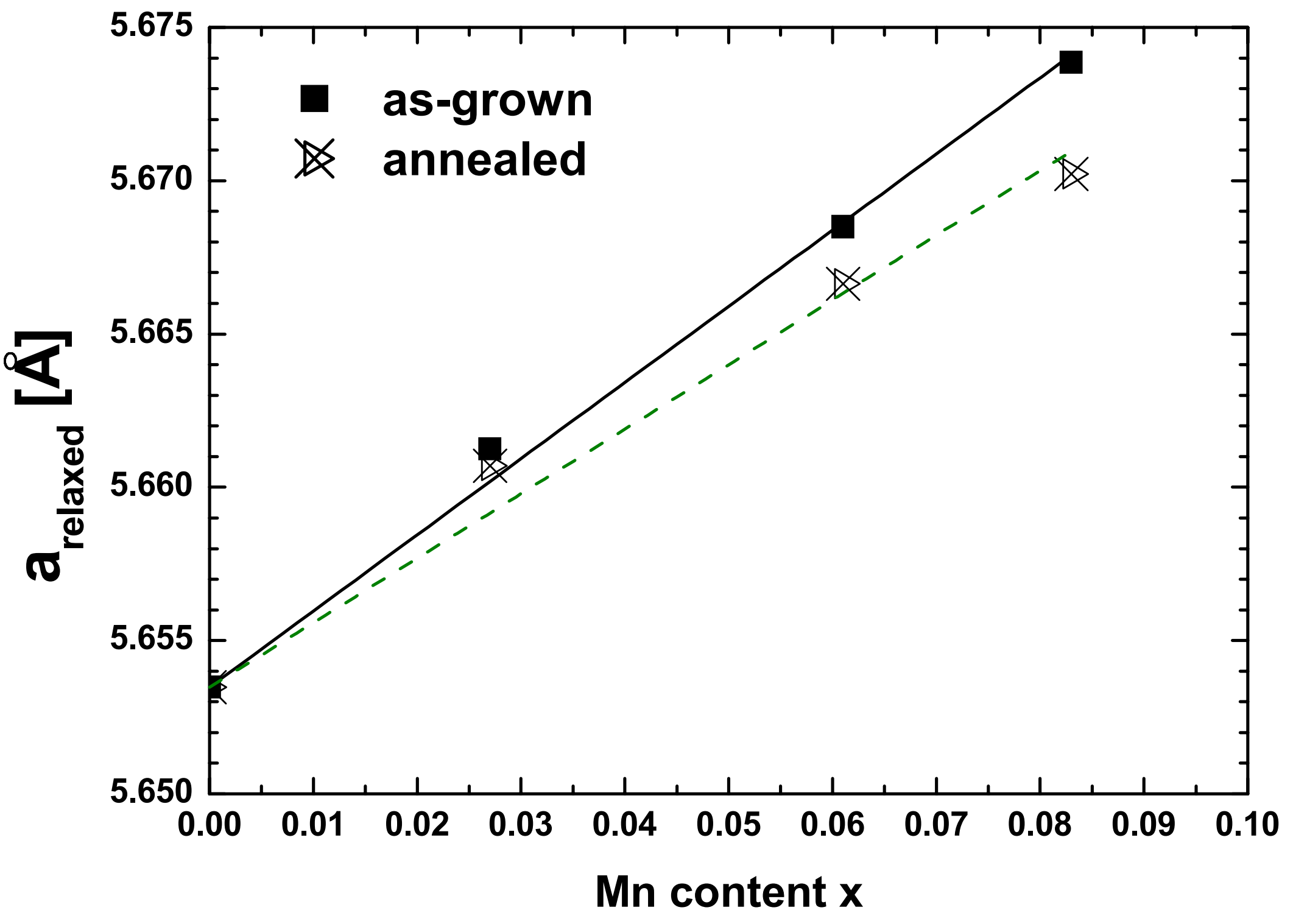

Fig. 6

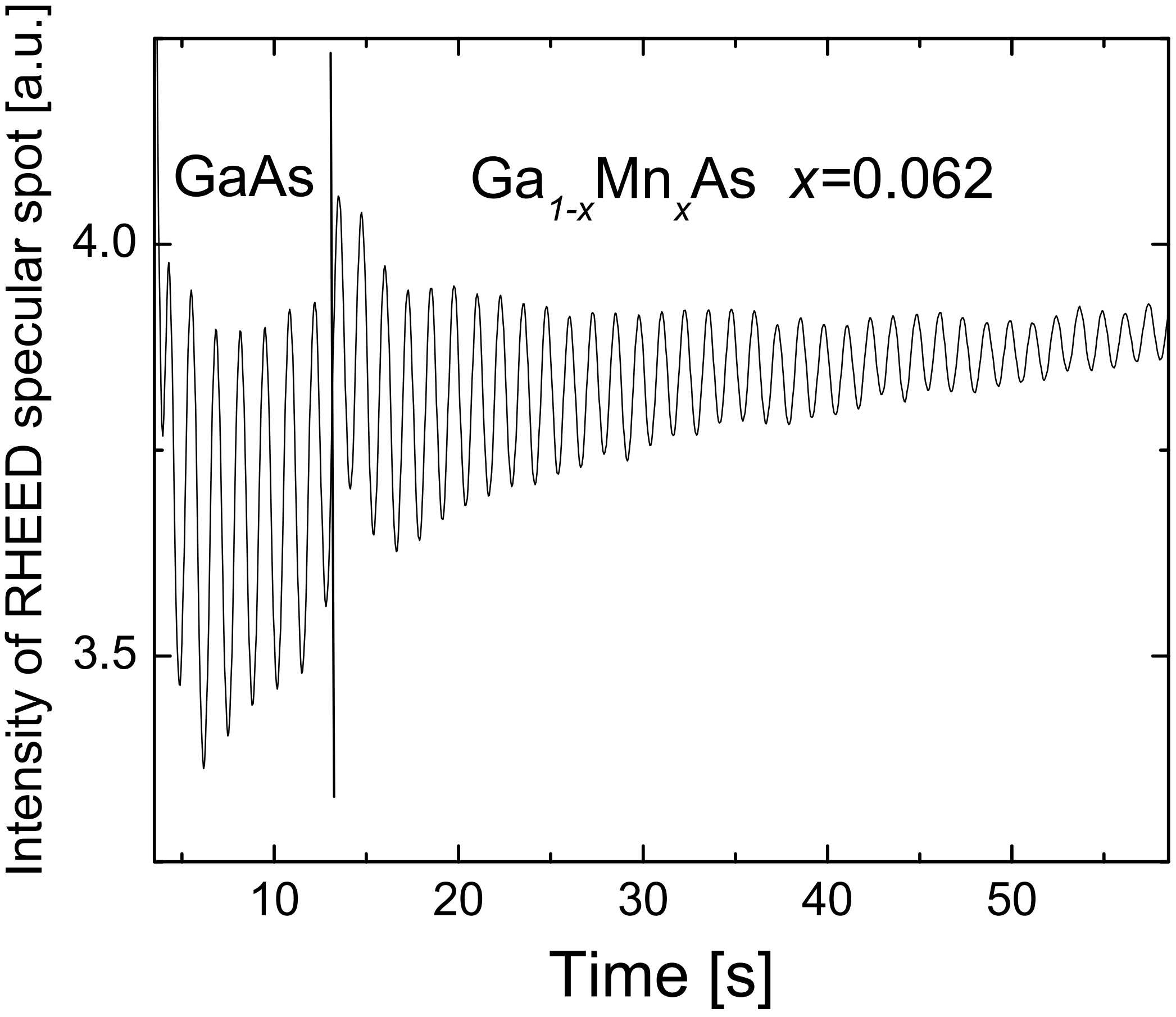

Fig. 7